\documentclass{article}
\usepackage{spconf,amsmath,graphicx}

\usepackage{booktabs}
\usepackage{multirow}
\usepackage{hyperref}

\usepackage{enumitem}
\setlist{nosep, leftmargin=14pt}

\usepackage{mwe} 

\title{RICAU-Net: Residual-block Inspired Coordinate Attention U-Net for Segmentation of Small and Sparse Calcium Lesions in Cardiac CT}

\usepackage{tikz}

\newcommand\copyrighttext{%
  \footnotesize \textcopyright 2025 IEEE. Personal use of this material is permitted. Permission from IEEE must be obtained for all other uses, in any current or future media, including reprinting/republishing this material for advertising or promotional purposes, creating new collective works, for resale or redistribution to servers or lists, or reuse of any copyrighted component of this work in other works.}

\newcommand\mycopyrightnotice{%
\begin{tikzpicture}[remember picture,overlay]
\node[anchor=south,yshift=10pt] at (current page.south) {\fbox{\parbox{\dimexpr0.9\textwidth-\fboxsep-\fboxrule\relax}{\copyrighttext}}};
\end{tikzpicture}
}

\name{Doyoung Park$^{1,2}$, Jinsoo Kim$^{3}$, Qi Chang$^{4}$, Shuang Leng$^{1,2}$, Liang Zhong$^{1,2}$, Lohendran Baskaran$^{1,2,\star}$\thanks{$^{\star}$Corresponding author.}}
\address{$^{1}$National Heart Centre Singapore, Singapore\\
$^{2}$CVS.AI, National Heart Research Institute of Singapore, Singapore\\
$^{3}$Independent Researcher, Republic of Korea\\
$^{4}$Department of Computer Science, Rutgers University, Piscataway, NJ, USA\\}

\begin{document}

\maketitle

\mycopyrightnotice
\begin{abstract}
The Agatston score, which is the sum of the calcification in the four main coronary arteries, has been widely used in the diagnosis of coronary artery disease (CAD). However, many studies have emphasized the importance of the vessel-specific Agatston score, as calcification in a specific vessel is significantly correlated with the occurrence of coronary heart disease (CHD). In this paper, we propose the Residual-block Inspired Coordinate Attention U-Net (RICAU-Net), which incorporates coordinate attention in two distinct manners and a customized combo loss function for lesion-specific coronary artery calcium (CAC) segmentation. This approach aims to tackle the high class-imbalance issue associated with small and sparse CAC lesions. Experimental results and the ablation study demonstrate that the proposed method outperforms the five other U-Net based methods used in medical applications, by achieving the highest per-lesion Dice scores across all four lesions.
\end{abstract}

\section{Introduction}
\label{sec:intro}

The Agatston score \cite{num3}, also known as coronary artery calcium (CAC) score, is widely used as a non-invasive method for prediction of coronary artery disease (CAD) by using non-contrast cardiac computed tomography (CT) \cite{num2, num4, num5}. The CAC score is calculated as the sum of calcification in the four main coronary arteries: the left main coronary artery (LM), left anterior descending artery (LAD), left circumflex artery (LCX), and right coronary artery (RCA). With the advancement of artificial intelligence in the medical field, deep learning algorithms have shown its capability of locating the calcium in the coronary arteries and automating the calculation of the CAC score with high precision \cite{num11, num13, num14}. Although the total Agatston score has been widely used for decades, many studies \cite{num11, num14, num16} emphasized the importance of the lesion-specific CAC scores. The authors in \cite{num21} concluded that the presence of CAC in RCA was significantly correlated with coronary heart disease (CHD) and cardiovascular disease (CVD). Therefore, the importance of the per-lesion Agatston score is as significant as the conventional Agatston score in predicting CHD and CVD. 

To calculate the per-lesion Agatston score, the lesion-specific CAC segmentation is a mandatory process. However, there are two limitations when facilitating the lesion-specific segmentation of CAC. The first factor to consider is the size and sparsity of the lesions. CAC is located in the coronary arteries; therefore, the size of CAC is relatively smaller than that of other organs such as the lung, kidney, or liver. The second issue is the high class-imbalance problem in the datasets. CAC in LM is the least common lesion among the four main coronary arteries because LM is the shortest coronary artery \cite{num25}. Therefore, the number of CT slices where CAC in LM appeared is the smallest, and the size of the LM lesion is also small. Many studies \cite{num11, num13, num14, num16} focused on lesion-specific segmentations, but \cite{num13, num14} combined CAC in LM and LAD as one lesion due to the challenge of distinguishing CAC in LM from CAC in LAD.

\begin{figure*}[htb]
  \centering
  \centerline{\includegraphics[width=17.0cm]{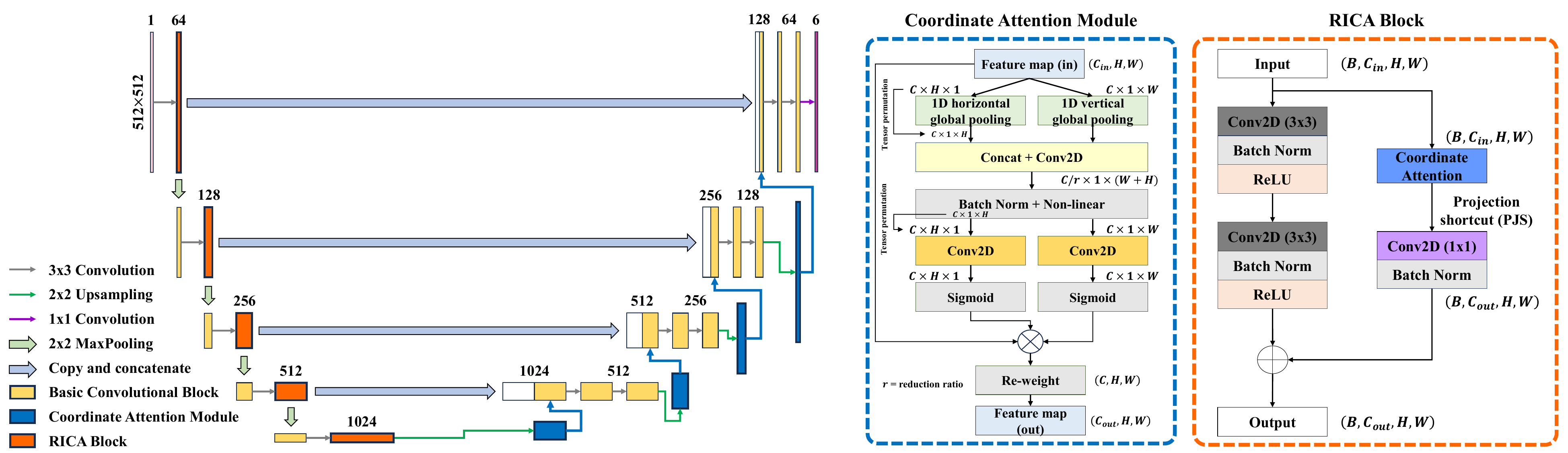}}
  \caption{Overview of the RICAU-Net architecture (left), and the CA module and the RICA block in RICAU-Net (right). The CA module illustrates the overall process of aggregating positional information from both horizontal and vertical directions of a feature map into channel attention. $r$ represents the reduction ratio, which is employed to decrease the complexity of the model by reducing the number of channels \cite{num27}. In our architecture, we utilized $r$ = 32.
  }
  \label{fig1}
\end{figure*}

Herein, we propose the Residual-block Inspired Coordinate Attention U-Net (RICAU-Net), an end-to-end deep learning algorithm for lesion-specific CAC segmentation using non-contrast cardiac CT scans. We also propose a customized combo loss function to tackle the segmentation of small and sparse lesions with a high class-imbalance problem. In the reporting deep learning algorithm, coordinate attention (CA) \cite{num27} is implemented as RICA blocks in the encoder and as CA modules in the decoder of the proposed model (Fig.~\ref{fig1}). This implementation aims to highlight the positional information of CAC as the locations of CAC were closely related to their surrounding vessels in CT images, which is substantial for the lesion-specific segmentation task. Through this study, we demonstrate the superior performance of the proposed algorithm and loss function compared to other U-Net based networks for lesion-specific CAC segmentation.

\section{Methods}
\label{sec:methods}

\subsection{Coordinate Attention Module}
\label{ssec:ca_module}

The CA module generates direction-aware and position-sensitive feature maps by breaking down channel attention into two 1D feature encoding processes that retain information in both horizontal and vertical directions \cite{num27}. These feature maps help the neural network accurately detect and locate regions of interest in the input image. Hence, the CA module was incorporated into the proposed architecture to emphasize the positional information of CAC and its surrounding arteries from the input images. The CA module is defined as:
\begin{align}
y_c = x \otimes a_w \otimes a_h
\end{align}
where $x$ and $y_c$ denote the input tensor and the output tensor of the CA module, $\otimes$ denotes element-wise multiplication, $a_w$ and $a_h$ denote width and height attentions in the CA module. As shown in Fig.~\ref{fig1}, a sigmoid activation is applied to the vertical and horizontal feature maps to generate width and height attention maps. Then, these attention maps are element-wise multiplied with the input tensor to generate the coordinate attention feature maps that have the same dimensions as the input tensor.

\subsection{Residual-block Inspired Coordinate Attention U-Net}
\label{ssec:RICAU}
The encoder of the RICAU-Net contains RICA blocks that utilize the CA information from the input in each layer. As illustrated in Fig.~\ref{fig1}, the design of the RICA block was inspired by the residual block in ResNet \cite{num37}, which consists of two convolutional layers and a skip connection. The skip connection includes a CA module and a projection shortcut (PJS) module. Since the dimensions of the feature maps from the CA module and the two convolutional layers are not equal, the PJS module is an essential step for matching the dimensions of the two groups of feature maps before the element-wise addition operation. The PJS module consists of $1\times1$ convolution followed by batch normalization. The RICA block is defined as:
\begin{align}
y_r = \mathcal{F}(x) \oplus (C_{pjs} \ast y_c)
\end{align}
where $y_r$ denotes the output tensor of a RICA block, $\mathcal{F}(x)$ denotes two convolutional layers; each convolutional layer followed by a batch normalization and a ReLU, $\oplus$ denotes element-wise addition, $C_{pjs}$ denotes a PJS module.

The decoder of the RICAU-Net is similar to that of the vanilla U-Net, but it incorporates four CA modules after each upsampling process. In this way, the network is capable of retaining the positional information of the upsampled features. The RICA block was not utilized in the decoder of the RICAU-Net because it could result in the loss of essential positional information when the feature maps from the CA module pass through the PJS module for dimension matching.

\subsection{Class-imbalance Problem and Loss Function}
\label{ssec:class-imbalance_and_loss}
\subsubsection{Class-imbalance Problem}
\label{sssec:class-imbalance}
Our non-contrast cardiac CT scan dataset shows a significant class imbalance between the negative samples (background, bone) and the positive samples (CAC in LM, LAD, LCX, RCA). For instance, Among training set, the smallest number of CAC was found in LM (1.3\%), whereas the largest number of CAC was observed in RCA (7.4\%). This phenomenon persisted in the validation and test sets as well. Among the validation and test sets, only 1.3\% and 0.4\% of images contained CAC in LM, respectively. It was more significant at the pixel level. In a $512\times512$ pixel image, one LM lesion could be as small and sparse as 5 pixels $(>1mm^2)$, whereas the remaining pixels would belong to the negative samples.

\subsubsection{Customized Combo Loss Function}
\label{sssec:loss}
Our dataset suffers from two major problems. The first issue is high class imbalance. The second issue is the small and sparse sizes of the target lesions. Focal loss \cite{num38} was first introduced to address the class imbalance problem in the object detection field. Exponential Logarithmic Loss (ELL) \cite{num41} was introduced to address the highly unbalanced object size problem in 3D segmentation. ELL is a type of combo loss that is based on a mixture of Cross-entropy loss and Dice loss. 

In this study, to overcome the two issues existing in our dataset, we propose a weighted Focal LogDice loss, which combined the weighted Focal loss and the Dice loss part of ELL with an exponential parameter. The weighted Focal LogDice loss is defined as:
\begin{align}
\mathcal{L}_{WFLogDL} = w_{Focal}\mathcal{L}_{WFL}+w_{Dice}\mathcal{L}_{ELDL}
\end{align}
where $\mathcal{L}_{WFLogDL}$ denotes the weighted Focal LogDice loss, $\mathcal{L}_{WFL}$ denotes weighted focal loss, $\mathcal{L}_{ELDL}$ denotes exponential logarithmic Dice loss,
$w_{Focal}$ and $w_{Dice}$ denote weights of each loss function, and $w_{Focal} = 0.4$ and $w_{Dice} = 0.6$ were used in this study.  

\begin{figure}[tb!]
  \centering
  \includegraphics[width=8.5cm]{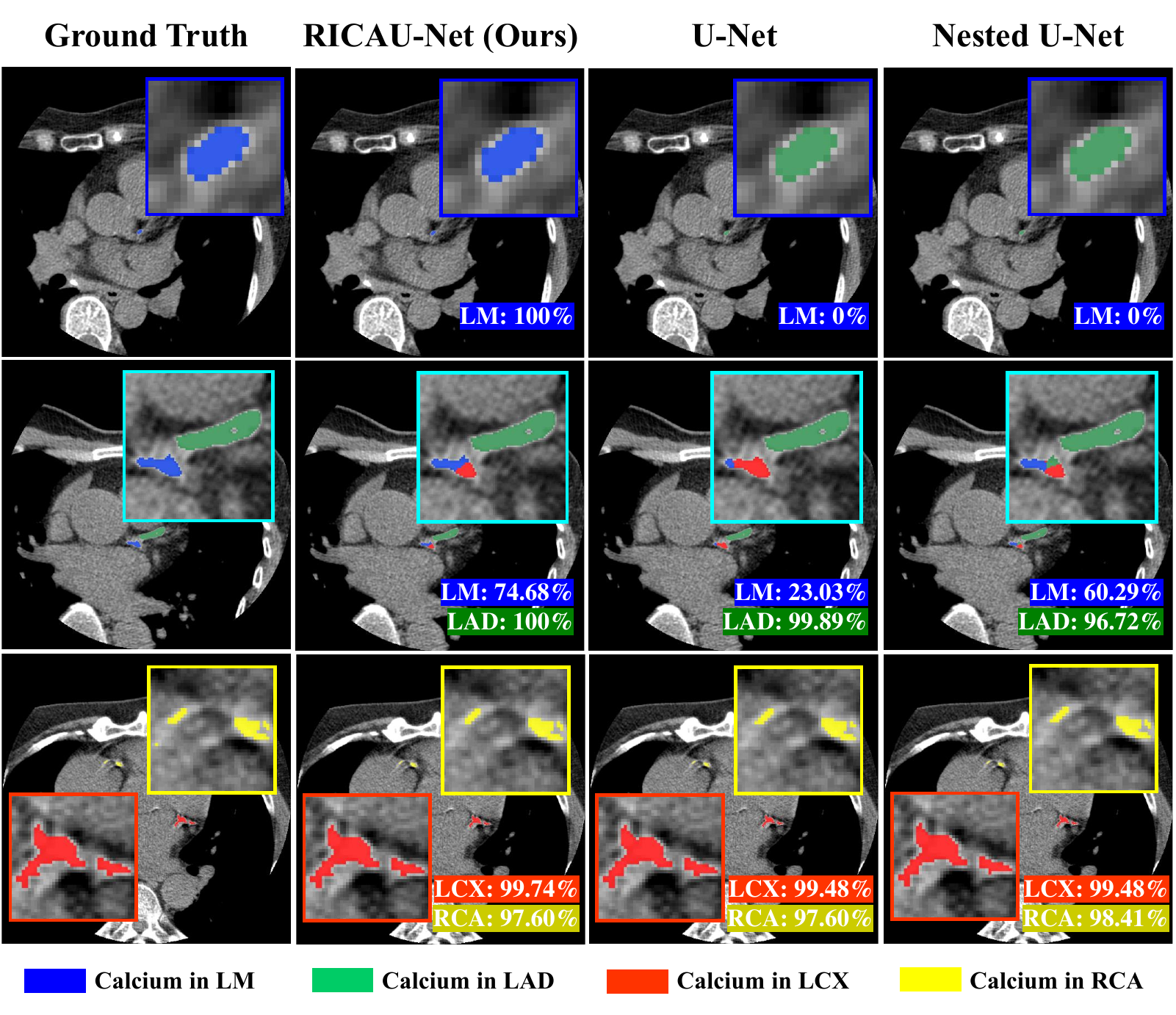}
  \caption{Visualization of the segmentation performance of RICAU-Net, U-Net, and Nested U-Net on three input images. The zoomed-in lesion images and corresponding per-lesion Dice scores with highlights are shown in the images. Our model significantly improves the segmentation performance for sparse and smaller CAC lesions, while also enhancing the results of other major ones.
  }
  \label{fig2}
\end{figure}

\section{Experiments and Results}
\label{sec:experiment-and-result}

\subsection{Dataset}
\label{ssec:dataset}
A total of 1,108 non-contrast cardiac CT scans were used in this study, acquired using four different scanners: SIEMENS, GE Medical Systems, PHILIPS, and TOSHIBA. The calcium lesions in the coronary arteries were manually annotated by five clinically trained specialists to create lesion-specific ground truth masks. The ground truth masks contained multiclass information on background, bone, CAC in LM, LAD, LCX, and RCA. A total of 612 CT scans from SIEMENS, GE Medical Systems, and Philips scanners were used for the training and validation sets. The training set (80\%) comprised 494 scans (484 subjects with CAC and 10 subjects without CAC), and the validation set (20\%) comprised 118 scans (112 subjects with CAC and 6 subjects without CAC). 496 CT scans from TOSHIBA scanner were used for the test set, comprising 278 subjects with CAC and 218 subjects without CAC.

\subsection{Preprocessing}
\label{ssec:preprocessing}
All the CT scans were sliced into 2D images with a resolution of $512\times512$ pixels. The raw Hounsfield Units (HU) of each slice were clipped to a range of [-150, 230], which represents the clinical CT windowing of the heart (window width: 380, window level: 40), followed by normalization to a range of floating-point numbers [0, 1]. In addition, random rotations of 5 and 10 degrees clockwise and counterclockwise, random center crops of $300\times300$ and $400\times400$ pixel resolutions, as well as the addition of Gaussian blur, Gaussian noise, and salt and pepper noise were employed as on-the-fly data augmentation methods to augment the training data.

\subsection{Implementation Details}
\label{ssec:implementation details}
In this study, a Pytorch container image version 23.06 (2.1.0a0+4136153) with Python 3.10.6 was used as the deep learning framework. All the experiments were conducted using four CPU cores (16GB) from AMD EPYC 7742 processor and two NVIDIA DGX-A100 graphics processing units (GPUs) with 40GB of RAM each. RICAU-Net was trained from scratch for 100 epochs with the weighted Focal LogDice loss function, a batch size of 16, the Adam optimizer with an initial learning rate of 1e-12, and the cosine annealing warm restart learning rate scheduler with a maximum learning rate of 1e-4, the first restart epochs of 50, the warmup restart epochs of 5, and the maximum learning rate scale factor of 0.5 at every restart \cite{num42, num43}.

\begin{table}[tb!]
  \caption{The per-lesion segmentation results of RICAU-Net and five other methods. All methods were trained with the weighted Focal LogDice loss function. The per-lesion Dice scores of the background and bone were above 99\% in all methods. The highest per-lesion Dice scores are in \textbf{bold}.
  }
  \label{tab1}
  \centering
  \begin{tabular}{ccccc}
    \toprule
    \multirow{2}{*}{Method} & \multicolumn{4}{c}{Per-lesion Dice Scores (\%)} \\  \cline{2-5} 
                            & LM         & LAD        & LCX       & RCA       \\  \midrule
    U-Net            & 53.02      & 93.62      & 84.13     & 91.37     \\
    Attention U-Net  & 52.27      & 93.00      & 83.83     & 90.90     \\ 
    Nested U-Net     & 57.50      & 93.42      & 82.50     & 91.87     \\ 
    VGG16 U-Net      & 52.94      & 91.82      & 82.93     & 86.99     \\
    TransUNet      & 51.40      & 91.57      & 83.86     & 89.99     \\ \midrule
    RICAU-Net (ours) & \textbf{59.35}      & \textbf{94.11}      & \textbf{87.03}     & \textbf{92.89}     \\          
    \bottomrule
  \end{tabular}
\end{table}

\subsection{Evaluation Metric}
\label{ssec:evaluation metric}
Dice score was used as the evaluation metric. It is one of the most commonly used evaluation metrics in semantic segmentation tasks because it measures the similarity between the ground truth and the model’s prediction. 

\subsection{Comparison with Other Methods}
\label{ssec:comparison}
The lesion-specific segmentation performance of RICAU-Net was evaluated by comparing it with five other methods: U-Net \cite{num28}, Nested U-Net \cite{num30}, Attention U-Net \cite{num32}, VGG16 U-Net, and TransUNet \cite{chen2024transunet}. All methods were trained using the same datasets and implementation details as RICAU-Net. For TransUNet, the pretrained model of R50+ViT-B\_16 on ImageNet21 was used for fine-tuning. The results are shown in Table~\ref{tab1}. The proposed method achieved the highest per-lesion Dice scores for all four lesions: 59.35 (LM), 94.11 (LAD), 87.03 (LCX), and 92.89 (RCA). For CAC in LM, considered as small and the scarcest lesions in our datasets, the top three methods with the highest LM Dice score were RICAU-Net, Nested U-Net, and U-Net. Among them, RICAU-Net outperformed Nested U-Net by 1.85\% and U-Net by 6.33\%. The segmentation performances of these methods were visualized in Fig.~\ref{fig2}. Overall, RICAU-Net demonstrated superior and more consistent results compared to other methods for lesion-specific segmentation, particularly for the rarest lesion (CAC in LM).

\subsection{Ablation Study}
\label{ssec:ablation}
An ablation study was conducted to validate the effectiveness of the weighted Focal LogDice loss in segmenting small and sparse lesions with high class imbalance. Four different loss functions were compared using RICAU-Net: Cross-entropy loss, weighted Focal loss, weighted Focal Dice loss, and weighted Focal LogDice loss. All the loss functions were trained using the identical implementation details as RICAU-Net. Per-lesion Dice scores of the four different loss functions with RICAU-Net are shown in Table~\ref{tab2}. RICAU-Net with the proposed loss function achieved the highest Dice scores for all four lesions, and a significant improvement in the Dice score for CAC in LM was observed. The proposed loss function showed a 10.51\% higher Dice score for CAC in LM compared to the lowest score. This demonstrated that the weighted Focal LogDice loss function is useful in the segmentation of small and sparse lesions, especially for classes experiencing a high class-imbalance problem.

\begin{table}[tb!]
  \caption{Per-lesion Dice scores of RICAU-Nets with four different loss functions. All the loss functions were weighted except for the cross-entropy loss. The highest per-lesion Dice scores are in \textbf{bold}.
  }
  \label{tab2}
  \centering
  \begin{tabular}{ccccc}
    \toprule
    \multirow{2}{*}{Loss Function} & \multicolumn{4}{c}{Per-lesion Dice Scores (\%)} \\  \cline{2-5} 
                            & LM         & LAD        & LCX       & RCA       \\  \midrule
    Cross-Entropy                 & 48.84      & 92.63      & 85.04     & 91.78     \\
    Focal                & 56.63      & 92.85      & 84.57     & 90.42     \\
    Focal Dice           & 55.48      & 93.68      & 85.00     & 91.42     \\ \midrule
    Focal LogDice (ours) & \textbf{59.35}      & \textbf{94.11}      & \textbf{87.03}     & \textbf{92.89}     \\ 
    \bottomrule
  \end{tabular}
\end{table}

\section{Conclusion}
\label{sec:conclusion}
In this study, we proposed a deep learning architecture called RICAU-Net and a customized combo loss function named the weighted Focal LogDice loss for the lesion-specific CAC segmentation using non-contrast cardiac CT scans that generally suffers from a high class-imbalance problem with small and sparse lesion datasets. When the proposed method was compared with other methods used in medical image segmentation tasks, our method exhibited the highest per-lesion Dice scores for all four lesions. The ablation study emphasized the importance the proposed loss function in segmenting small and sparse lesions, particularly for CAC in LM which was small and the scarcest in our dataset. This algorithm can be further utilized to streamline the manual process of identifying lesion-specific calcium plaques and calculating lesion-specific calcium scores by cardiologists.

\section{Compliance with ethical standards}
\label{sec:ethics}
This research study was conducted retrospectively using human subject data. Ethical approval was received from the Institutional Review Board of SingHealth (IRB 2022-2521).



\section{Acknowledgments}
\label{sec:acknowledgments}
\begin{sloppypar}
This work was supported by National Medical Research Council (NMRC) of Singapore (grant numbers: NMRC/CG2/001a/2021-NHCS, CG21APR1006, TA21nov-000), SingHealth-Duke-NUS ACP Philanthropy grant (HRDUK230600) and SingHealth-Duke-NUS ACP NCRS grant (07/FY2021/P2/12-A93). L.Z. is supported by NMRC under its Clinician-Innovator Award Senior Investigator (CIASI24jan-0001) and Health Service Research Award (MOH-000358), SingHealth Duke-NUS Academic Medical Centre AM strategic fund award (07 FY2023 HTP P2 15-A2) and Industry Alignment Fund Pre-Positioning (IAFPP award) (H20c6a0035). The computational work for this article was fully performed on resources of the National Supercomputing Centre (NSCC), Singapore (https://www.nscc.sg).
\end{sloppypar}

\bibliographystyle{IEEEbib}
{\small\bibliography{main}}
\end{document}